%% file: main.tex
\PassOptionsToPackage{table}{xcolor}
\documentclass[acmsmall,screen,nonacm]{acmart}

\usepackage{booktabs,tabularx}
\usepackage{graphicx}
\usepackage{url}

\usepackage{multirow}

\usepackage{threeparttable}

\newcommand{\system}{\textsc{SymDec}}

\usepackage{tcolorbox}

\newtcolorbox{rqanswer}[1]{
    colback=black!3,
    colframe=black!45,
    boxrule=0.4pt,
    arc=0pt,
    left=6pt,
    right=6pt,
    top=5pt,
    bottom=5pt,
    before skip=8pt,
    after skip=8pt,
    fonttitle=\bfseries,
    colbacktitle=black!3,
    coltitle=black,
    title={Answer to #1}
}

\usepackage{tikz}
\usepackage{listings}
\usetikzlibrary{arrows.meta}

\definecolor{caseblue}{RGB}{35,76,112}
\definecolor{casered}{RGB}{165,45,45}
\definecolor{casegreen}{RGB}{30,110,65}

\lstdefinestyle{motivationcode}{
    language=C,
    basicstyle=\ttfamily\scriptsize,
    keywordstyle=\color{caseblue},
    commentstyle=\color{black!55},
    columns=fullflexible,
    keepspaces=true,
    showstringspaces=false,
    breaklines=true,
    aboveskip=3pt,
    belowskip=3pt
}

\begin{document}

\title{Revisiting Re-executability: Coverage-Guided Symbolic Feedback to Improve Decompilation}

\author{Yuxin Hu}
\authornote{Both authors contributed equally to this research.}
\affiliation{%
  \institution{Vanderbilt University}
  \city{Nashville}
  \state{TN}
  \country{USA}}

\author{Xiaohan Wang}
\authornotemark[1]
\affiliation{%
  \institution{Vanderbilt University}
  \city{Nashville}
  \state{TN}
  \country{USA}}

\author{Kevin Leach}
\affiliation{%
  \institution{Vanderbilt University}
  \city{Nashville}
  \state{TN}
  \country{USA}}


\input{sections/abstract}

\maketitle

\input{sections/01_introduction}

\input{sections/03_motivation_symbolic_feedback}
\input{sections/04_system_design}

\input{sections/05_implementation}

\input{sections/06_evaluation_results}

\input{sections/07_discussion_limitations}

\input{sections/08_related_work}

\input{sections/09_conclusion}

\input{sections/references}

\end{document}

%% file: sections/abstract.tex
\begin{abstract}
When source code is unavailable, decompilation provides
a source-level representation of a binary.
However, code recovered by conventional or LLM-assisted
decompilation may not preserve the binary's original
behavior.
Recent state-of-the-art approaches evaluate
re-executability using fixed test suites, but passing
these suites does not ensure that the recovered code
preserves the original behavior.
We present \system{}, an LLM-assisted decompiled-code
repair system that uses the original binary as a
behavioral reference.
We use coverage-guided symbolic execution to explore
the I/O-level behavior of the original and recompiled
binaries, and symbolic comparison to identify inputs
that expose behavioral differences.
These inputs are executed on both binaries to confirm
the differences, which are returned as structured
feedback to guide subsequent LLM revisions, producing re-executable candidates that more closely match the behavior of the original binary.

We evaluate \system{} on 120 functions from ExeBench
and DecompileBench, compiled for x86-64 and ARM64
at optimization levels O0--O3, yielding 960 binary
decompilation tasks.
More than half of the tasks yield candidates that
pass the supplied test suite and complete symbolic
comparison without detected behavioral differences,
providing evidence of behavioral consistency beyond
test-suite re-executability alone.
Among candidates that pass all supplied tests, symbolic comparison exposes replay-confirmed behavioral differences in 4.5\% of cases, showing that test-suite re-executability can miss behavioral discrepancies.
Under the same input-generation budget, symbolic
execution achieves higher mean line and branch
coverage than black-box fuzzing, demonstrating
more effective exploration of the evaluated functions.
Although test-suite re-executability alone does not
establish behavioral equivalence, symbolic feedback
also improves this conventional metric, increasing
the mean single-step test-pass rate from 90.1\%
to 93.3\%.
These results demonstrate that symbolic comparison
provides stronger evidence of behavioral consistency
than test-suite re-executability alone and supplies
effective feedback for repairing decompiled code.
\end{abstract}

%% file: sections/01_introduction.tex
\section{Introduction}

Decompilation helps developers understand and maintain software when its source code is unavailable.
When decompiled code is used as an executable replacement,
it must compile and preserve the original binary's
observable behavior~\cite{wong2025decllm,zou2024dhelix}.
Recent LLM-assisted approaches improve the readability and re-executability of decompiled code~\cite{hu2024degpt,tan2024llm4decompile,liu2025codeinverter,zhang2026code}.
However, prior work evaluates re-executability
by checking whether decompiled code passes
the supplied tests, including I/O examples in
ExeBench and test assertions in
HumanEval-Decompile~\cite{tan2024llm4decompile}.
Passing these tests establishes agreement only
on the tested inputs, leaving potential behavioral
differences on other untested inputs undetected.
Consequently, a candidate considered re-executable
under this evaluation may still fail to preserve
the original binary's behavior.
The original binary provides a behavioral reference that we can use to generate additional feedback and guidance during decompilation.
In this paper, we study how symbolic comparison can expose
behavioral differences missed by existing
test suites  and  guide LLM-assisted repair\footnote{In this paper,
\emph{repair} refers to modifying decompiled source code
to reproduce the original binary's behavior,
rather than correcting defects that may be present in the original
program.} of decompiled code.

\begin{figure}[t]
    \centering
    \includegraphics[width=0.85\linewidth]{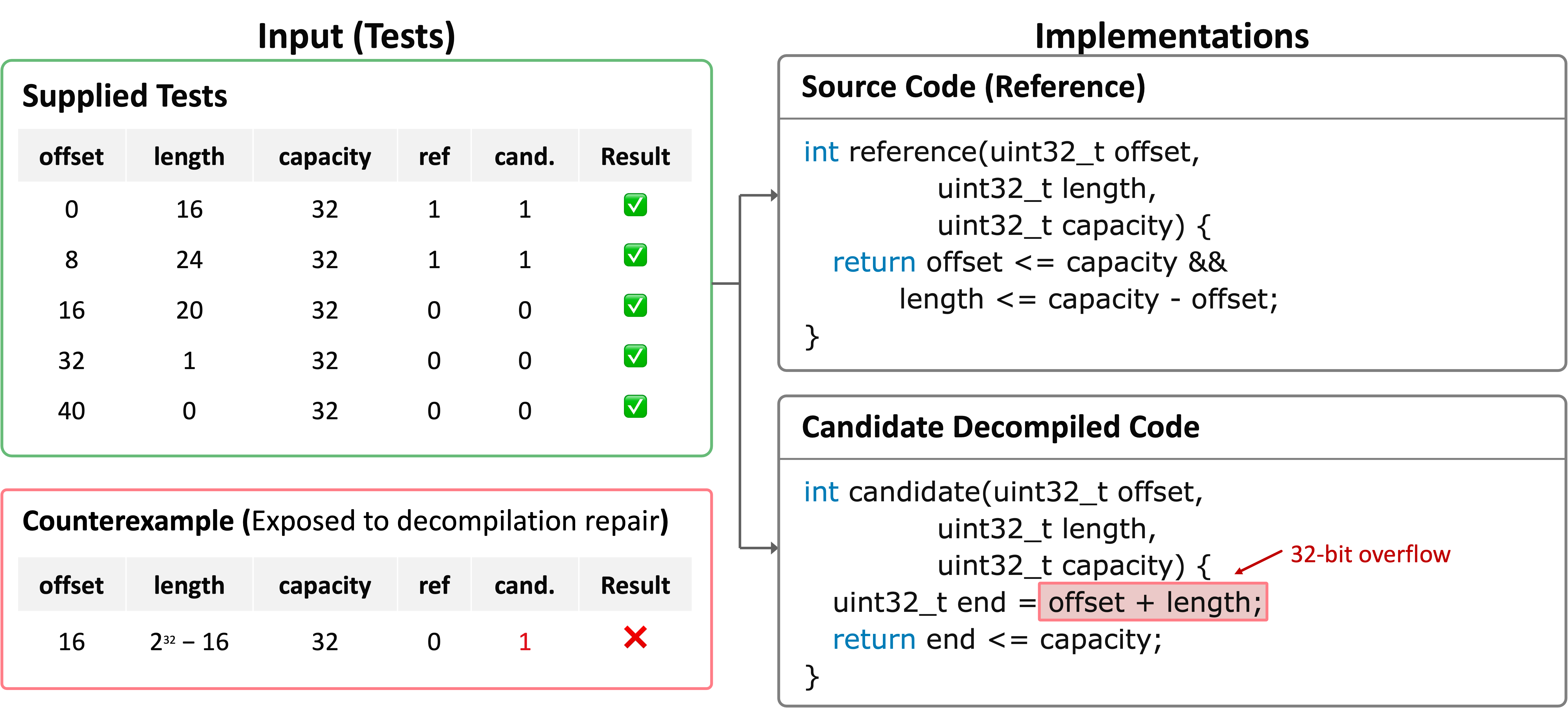}
    \caption{A motivating example of a buffer-range check. The candidate decompilation matches the reference on every supplied test, including boundary cases, yet is not semantically equivalent.
    Symbolic comparison searches for inputs that expose this difference.
    On the counterexample, the 32-bit sum $\mathit{offset}+\mathit{length}$ wraps to $0$, so the candidate accepts an out-of-range request that the reference rejects.}
    \label{fig:motivation}
\end{figure}

Consider a motivating example of a buffer-range check in~\autoref{fig:motivation}.
The original function rejects requests that extend beyond the buffer.
A candidate instead adds the offset and length using 32-bit unsigned arithmetic and compares the sum with the buffer's capacity.
The two agree unless the addition wraps around.
A wrapped sum can make the candidate accept an out-of-range request.
The supplied test suite may miss this case and
therefore provide no feedback about the behavioral
mismatch.
Symbolic comparison can identify an input that
exposes this mismatch.
Executing both binaries on that input confirms
the difference and provides the LLM with a
concrete counterexample to guide its next revision.

Prior work uses differential analysis to improve
decompilation. CHISEL uses differential fuzzing against
the original binary to guide iterative repair
\cite{chisel2026}. D-Helix compares original and
recompiled binaries symbolically and adjusts
decompiler heuristics to correct detected errors
\cite{zou2024dhelix}.
Our focus is test-guided symbolic comparison that
provides feedback to an LLM editing the candidate
decompiled source. This requires searching beyond
the supplied tests within a limited exploration
budget and turning detected differences into
concrete feedback for the next revision.
We evaluate both the behavioral differences detected
by symbolic comparison beyond the supplied tests
and the effect of symbolic feedback on decompiled-code
repair.

Symbolic execution tracks how program behavior depends
on input values and uses path constraints to generate
concrete inputs.
Branches and loops can make exhaustive exploration
impractical. Prior work uses code coverage and
execution-path information to prioritize exploration
and improve test
generation~\cite{cadar2008klee,seo2014we,yao2025empc}.
In our setting, traces from the supplied tests help
prioritize code not yet covered, but previously
covered code can still behave differently on other
inputs. We therefore combine coverage priority
with execution-history diversity to explore
alternative inputs and paths through visited code.
Because decompilation and recompilation can change
control flow, we explore the two binaries independently
and compare their outputs under shared inputs.
When symbolic comparison finds an input for which
the modeled outputs differ, we execute both binaries
on that input to check the difference.
The counterexample input and confirmed output
difference then provide concrete feedback for
the LLM's next revision.

Therefore,
in this paper, we present \system{}, a framework that uses symbolic comparison to expose differences between the original and recompiled binaries and to guide an LLM in repairing decompiled code.
We use an LLM to propose source edits, while a
controller component
coordinates compilation, testing, and symbolic comparison.
Our framework explores the original and recompiled
binaries independently to obtain path constraints
and symbolic outputs for comparison.
The comparison checks return values and modeled output
buffers under shared inputs without requiring matching
control flow.
The controller requires a comparison result for the current candidate before acceptance; confirmed differences prompt further repair.
The comparison uses the supplied function interface
and models of memory and supported external functions,
with exploration bounded by configured resource budgets.
By checking candidate behavior beyond the supplied
test suite and confirming detected differences through
execution, \system{} provides additional evidence of
behavioral consistency and concrete feedback for repair.

We evaluate \system{} on 120 functions from ExeBench and DecompileBench,
compiled for x86-64 and ARM64 at O0--O3 to form 960 binary decompilation tasks.
We evaluate difference detection and repair feedback separately.
For each repair run, we select the first candidate
that passes all ten supplied tests (IO10).
Among these candidates, symbolic comparison detects
replay-confirmed differences in 4.5\%, compared
with 2.4\% for black-box fuzzing.
These differences expose behavioral mismatches
that test-suite-based re-executability leaves undetected.
For repair, a paired single-step ablation starts
each feedback group from the same candidates
that compile but fail IO10.
Although passing IO10 alone does not establish
behavioral equivalence, symbolic feedback also
improves this conventional metric.
Over three trials,
it achieves a mean
single-step IO10 pass rate of 93.3\%, compared
with 90.1\% for I/O feedback alone and 90.7\%
for I/O feedback augmented with fuzzing.
Together, these results show that symbolic comparison
extends behavioral assessment beyond
the current state-of-the-art use of fixed test suites
while providing feedback that improves the overall quality of
decompiled code.

Our contributions are as follows:
\begin{enumerate}
    \item An approach that combines testing with coverage-guided symbolic execution to provide stronger assurance of semantic consistency between the original and recompiled binaries.

    \item \system{},
    an LLM-assisted decompilation framework that uses inputs that expose behavioral differences between the original and recompiled binaries to guide decompilation revisions,
    and reassesses these revised candidates through testing and symbolic comparison.

    \item An evaluation across 960 binary decompilation tasks showing that symbolic execution achieves higher average coverage than black-box fuzzing and detects behavioral differences not exposed by given tests.
    Our ablation study shows that adding symbolic feedback increases the fraction of decompiled candidates that pass all supplied tests after one revision.
\end{enumerate}

%% file: sections/03_motivation_symbolic_feedback.tex
\section{Motivation: Why Symbolic Feedback?}
\label{sec:motivation}
Below, we discuss test-based re-executability and symbolic feedback for iterative decompilation.

\subsection{The Gap Between Test-Based Re-executability and Semantic Equivalence}

Recent decompilation work evaluates recovered code through execution against benchmark tests.
LLM4Decompile~\cite{tan2024llm4decompile} and CodeInverter~\cite{liu2025codeinverter} report re-executability on ExeBench by checking whether reconstructed code executes successfully and satisfies the
given test suite of test cases per sample.
Compilation makes the candidate executable within that limited evaluation harness;
matching the expected outputs is then used as evidence to conclude behavioral agreement.
This procedure supports reproducible comparisons between decompilers,
but ultimately does not provide rigorous assessment of semantic equivalence or similarity.

The distinction is between agreement on a finite test set and agreement throughout the intended input domain under the same observable-behavior definition.
SLaDe~\cite{armengol2024slade} separates these notions,
using finite-set I/O accuracy as a practical evaluation criterion while noting that exhaustive exploration of a finite domain is needed to guarantee equivalence through testing.
ExeBench~\cite{10.1145/3520312.3534867} distinguishes simple I/O,
whose observed behavior is constant or identity,
from richer I/O that rules out these trivial implementations.
The discussion from ExeBench illustrates how tests can underconstrain an implementation:
a simple return value can satisfy some test sets without establishing recovery of the original computation.
Even when program behavior is observable through straightforward I/O, richer test suites do not necessarily establish a complete behavioral specification.
DecompileBench~\cite{gao2025decompilebenchcomprehensivebenchmarkevaluating}
further highlights that input--output checks can miss behaviors
such as global-state changes and heap manipulation.
This motivates using the original binary as an executable reference to search for behavioral differences beyond the supplied tests.

\subsection{Symbolic Feedback for Iterative Decompilation}

Existing systems already use validation feedback to revise decompiler output.
MCGD~\cite{zhang2026mcgd} checks syntax,
compilation,
and execution in successive stages,
returning failing inputs and expected versus actual outputs to specialized repair agents.
CHISEL~\cite{chisel2026} combines compiler diagnostics with type-aware,
coverage-guided differential fuzzing against the original binary,
providing repair feedback without requiring a supplied test suite.
These approaches establish the value of checking and revising generated candidates.
They also show that feedback need not be restricted to fixed tests:
dynamic input generation can expose additional discrepancies during repair.
The remaining question is what additional behavioral evidence symbolic analysis can contribute to this process.

Symbolic comparison searches for distinguishing inputs through path constraints and conditions on differing outputs.
Differential Symbolic Execution~\cite{person2008dse} uses symbolic reasoning to characterize behavioral changes,
while D-Helix~\cite{zou2024dhelix} applies symbolic comparison to original and recompiled binaries for decompiler testing.
In our setting,
a proposed distinguishing input is replayed on both binaries;
the confirmed input and differing outputs then provide concrete feedback for source revision.
This constraint-based search offers a complementary way to obtain feedback after the supplied tests pass.
Its usefulness relative to fuzzing depends on the programs and analysis budget,
which motivates evaluating both difference detection and improvements in decompilation quality.

Each revision changes the candidate being evaluated,
so agreement on previously checked inputs must be reassessed and further differences sought in the updated candidate.
The objective is to improve the behavioral fidelity of the recovered source to the original binary.
The evidence remains subject to execution models,
path exploration,
and solver resources:
replay confirms individual behavioral differences,
while an absence of discovered differences does not establish full semantic equivalence.

%% file: sections/04_system_design.tex
\begin{figure}[t]
\centering
\includegraphics[width=\linewidth]{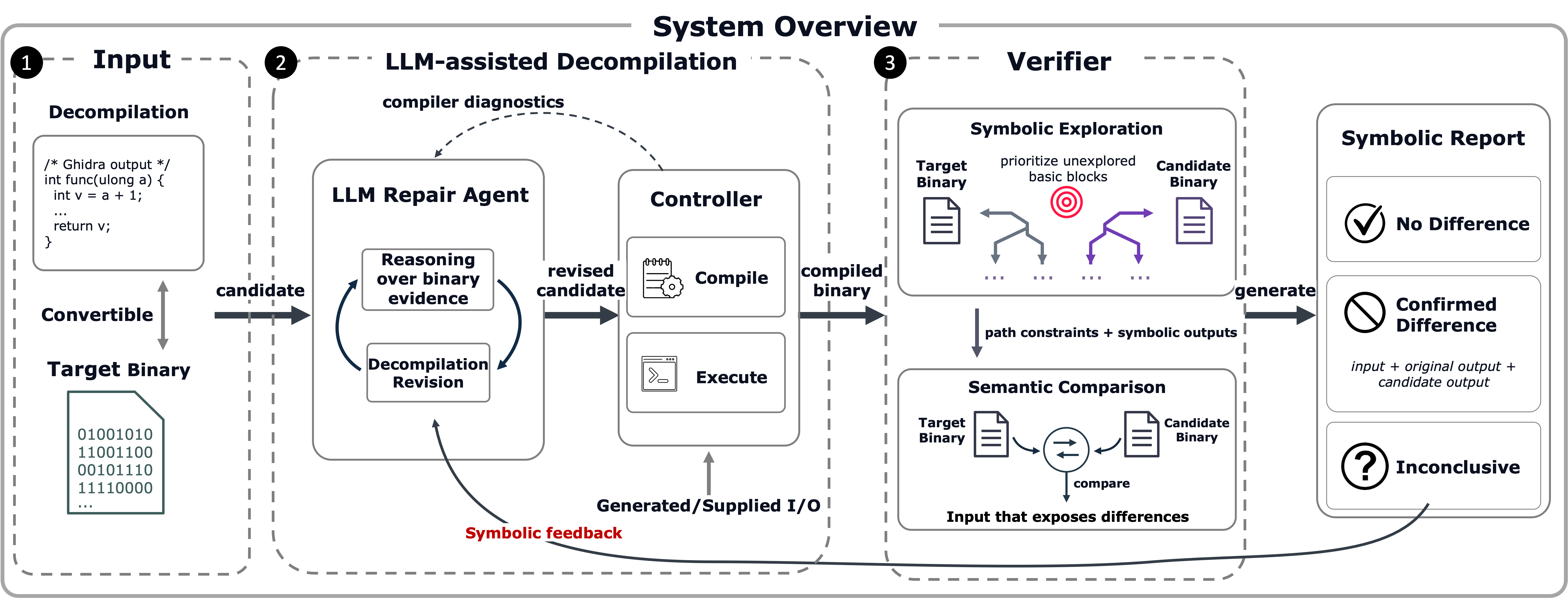}
\caption{
System overview.
(1) The output of a decompiler and the target binary serve as input.
(2) The repair agent iteratively revises the decompiled
code, guided by compiler diagnostics and execution results.
(3) The verifier compares the recompiled binary
and target binary and reports No Difference,
Confirmed Difference, or Inconclusive. The report is fed back to the repair agent as symbolic feedback for the next round of revision.
The final output is revised decompiled source that, once compiled, is more likely to be semantically equivalent to the original input binary.
}
\label{fig:overview}
\end{figure}

\section{System Design}
\label{sec:design}
\label{sec:design-overview}

\system{} repairs decompiled code to reproduce the
observable behavior of the original binary.
As illustrated in Figure~\ref{fig:overview},
\system{} consists of three components: a repair agent, a controller, and a verifier.
We start with a snippet of decompiled code, an input binary, and a fixed set of test I/O pairs.
The repair agent follows a ReAct-style interaction
pattern~\cite{yao2022react} to examine the current
decompiled candidate and available feedback,
invoke symbolic comparison, and propose code revisions.
The controller executes the repair agent's actions,
coordinates compilation and the supplied I/O tests,
manages candidate revisions, and enforces the repair
budget and acceptance criteria.
The verifier performs symbolic comparison between
the original and recompiled binaries and checks
potential counterexamples through concrete replay.
It returns the comparison results as feedback
to the repair agent.

The components provide complementary feedback for repair.
Compiler diagnostics identify errors that prevent
compilation, and concrete tests check agreement with
the original binary on the supplied inputs.
Symbolic comparison extends these checks by searching
for additional inputs that expose behavioral differences.
The repair agent uses this feedback to revise the
candidate's source code, which the controller
then recompiles.
The verifier compares the resulting executable with
the original binary by examining their observable
outputs under the same inputs.
This comparison does not require matching instructions
or control-flow structures, allowing the candidate
to reproduce the original behavior through a different
organization of the code.
\system{} aims to produce a revision of the decompiled
code that (1) recompiles successfully, and (2) when
executed, matches the I/O-level behavior of the
original input binary.

\subsection{Repair Workflow}
\label{sec:design-workflow}

The repair process begins with Ghidra's decompiled
output~\cite{ghidra}---the initial decompiled
\emph{candidate}.
The controller attempts to compile the candidate.
If compilation fails, it returns compiler errors
to the repair agent to guide
revision~\cite{liu2026autodecompiler}.
Once compilation succeeds, the controller runs
the configured I/O tests, compares the candidate's
outputs with the expected outputs, and returns
the test results to the repair agent.
For a failing test, the controller returns the failing input and differing outputs to the repair agent to examine the discrepancy and propose a revision.

The repair agent uses the available feedback to choose
between revising the code and requesting further
analysis.
It can obtain additional binary context to inspect
the original implementation or invoke symbolic
comparison to search for behavioral differences
beyond the given tests.
When concrete replay confirms a counterexample (i.e., an input whose corresponding output differs between the decompiled candidate and original binary),
the verifier returns the input and the outputs
observed on both binaries.
The repair agent uses this evidence to guide a
source-level repair.
After each attempted repair, the controller recompiles the
candidate and reruns the concrete tests, returning
the updated results for the next repair step.

Once the candidate compiles and passes the given tests,
the controller invokes the verifier for a symbolic
assessment.
It uses this assessment to
determine if any further behavioral differences exist between the decompiled candidate and input binary.
If a behavioral difference is confirmed, the
controller returns the counterexample (i.e., a test input obtained from symbolic assessment) to the repair agent
to guide the next revision.
The repair loop ends when the controller accepts
a candidate or the repair budget (i.e., time) is exhausted.
The final candidate is returned together with
its compilation, test, and symbolic-comparison
results.

\subsection{Coverage-Guided Symbolic Exploration}
\label{sec:impl-coverage}

The verifier symbolically executes the original
binary and the recompiled candidate to obtain
execution paths and symbolic outputs for comparison.
Both binaries use a shared modeled input domain
defined by the supplied function interface and
the buffer sizes specified in the input model.
Scalar arguments are represented by symbolic values,
while pointer arguments refer to finite modeled
buffers.
The two binaries are explored independently because
decompilation and recompilation can change their
branch structures and basic-block boundaries.
This exploration seeks behavior beyond the supplied
tests, providing a broader basis for assessing
whether the candidate preserves the original
binary's behavior.

When a branch has multiple feasible outcomes,
symbolic execution creates a separate state for
each outcome.
As branching continues, pending states can accumulate,
leaving many paths to explore within a limited
resource budget.
The order of exploration therefore affects which
paths can be examined before the budget is exhausted.
Prior work uses coverage-oriented heuristics,
execution context, and path-cover information
to prioritize symbolic
exploration~\cite{cadar2008klee,seo2014we,yao2025empc}.
Our scheduler combines basic-block coverage priority
with execution-history diversity, using available
test traces to initialize its coverage information.
The visited-block set is updated as exploration
proceeds.
The scheduler alternates between preferring states
at unvisited blocks and selecting states according
to the distinctiveness of their execution histories.
The first priority directs exploration toward
code not yet exercised.
The second favors different paths, including
paths through previously visited blocks that
may carry different constraints or intermediate values.
States not selected remain available for later
exploration, and the scheduler periodically advances
the longest-waiting state to reduce repeated delays.

These priorities change exploration order without
restricting symbolic inputs to the supplied test
values.
Exploration returns completed paths and their
symbolic outputs, together with the exploration
status, for subsequent comparison.

\subsection{Semantic Consistency Assessment}
\label{sec:impl-verification}

The verifier uses the paths obtained from symbolic
exploration to compare the behavior of the two
binaries, following the general approach of symbolic
crosschecking~\cite{ramos2011practical,collingbourne2014symbolic}.
Each path supplies a condition describing the inputs
that follow it and symbolic expressions for its
return value and modeled memory outputs.
To compare a path from each binary, the verifier
asks the solver for an input that satisfies both
path conditions and makes at least one output differ.
This compares the binaries on the same input
without requiring their instructions or branch
structures to match.

If the solver finds such an input, it provides
a potential counterexample.
The verifier reconstructs the input and replays
it on both binaries, comparing their observed
return values and modeled memory outputs.
A mismatch reproduced by concrete replay is
reported as a confirmed difference.
The counterexample input and differing outputs
are then returned to the repair agent to guide
its next revision.

The verifier reports \textit{No diff} when exploration
and all required comparisons complete without
finding a difference within the modeled domain.
This provides stronger assurance of behavioral
consistency than agreement on the supplied tests alone.
\textit{Partial no diff} indicates that no difference
was found in the completed comparisons, but the
assessment remains incomplete, for example because
a resource limit was reached.
Further analysis may still expose a behavioral
difference.

\subsection{Feedback for Code Repair}
\label{sec:design-feedback}

The verifier converts comparison results into
structured feedback for the repair agent.
For a confirmed difference, the report includes
the counterexample input, the outputs observed
on both binaries, and the values that differ.
It identifies whether the discrepancy concerns
the return value or values written to an output
buffer through pointer arguments.
For example, two executions may return the same
status but produce different output arrays.
Inputs and outputs are presented with their types,
while preserving the exact values needed for replay.
The report also records the modeled scope,
comparison outcome, and completion status.

The repair agent examines this report together with
the current candidate and compiler or test feedback
to decide its next action.
A confirmed counterexample gives the repair agent
a reproducible input and the reference output
that the candidate should produce.
The repair agent can inspect the relevant code,
request additional binary context, or propose
a repair to address the discrepancy.
The feedback describes the behavior to correct
rather than a source-level change.
After each revision, the controller repeats
compilation and the supplied I/O tests and
coordinates further symbolic assessment following
the workflow in Section~\ref{sec:design-workflow}.
These checks determine whether the reported
difference remains and whether other failures
are detected.

By searching beyond the supplied test suite and
using confirmed counterexamples to guide revisions,
\system{} addresses behavioral differences that
test-based repair can leave undetected.
The resulting decompiled candidate is assessed
through both the supplied tests and symbolic
comparison, providing stronger evidence of
behavioral consistency than test-suite agreement
alone.

%% file: sections/05_implementation.tex
\section{Implementation}
\label{sec:implementation}

We apply \system{} to
x86-64 and ARM64 binaries.
This section describes cross-architecture execution,
candidate preparation, solver integration,
and counterexample replay within the repair loop described in Section~\ref{sec:design-overview}.

\subsection{Cross-Architecture Execution}
\label{sec:implementation-execution}

\system{} expects an input binary, a target function's entry address and calling interface, an initial decompilation candidate produced by Ghidra, and a set of I/O test pairs for that function.
The entry address identifies the original function
without requiring its name to be preserved
in the binary.
The given I/O pairs are used for concrete
testing, and their execution traces guide symbolic
exploration.
Candidates are compiled for the same architecture
as the original binary.
The verifier uses angr~\cite{shoshitaishvili2016state}
to initialize symbolic arguments according to the
given calling interface and place them in the
registers or stack locations required by the
target architecture's calling convention.
For concrete replay, x86-64 binaries run natively,
while ARM64 binaries run through QEMU user-mode
emulation~\cite{bellard2005qemu}.
Symbolic crosschecking compares implementations
under the same inputs~\cite{ramos2011practical}.
In our implementation, the original and candidate
binaries receive identical scalar values and
initial buffer contents.
Each binary uses separately allocated buffers,
so their addresses need not match.
During concrete replay, the verifier prepares
these inputs and collects the return value and
values written to output buffers.

During repair, the repair agent can request additional
binary context through the Ghidra
integration~\cite{ghidra}, including disassembly,
control-flow information, and referenced data.
This context helps the repair agent inspect the
original implementation when deciding how to
revise the candidate.

\subsection{Candidate Preparation and Input Layout}
\label{sec:implementation-adaptation}

Before a candidate can be tested or compared,
the implementation prepares it for compilation
and generates code to invoke the target function
using the given calling interface.
The compilation environment provides compatibility
definitions for Ghidra pseudo-types and helper
operations, such as \texttt{undefined4} for fixed-width
values and \texttt{CONCAT11} for byte concatenation.
Generating this calling code also requires identifying
the target function, since the repair agent may
introduce helper functions during repair.
The implementation permits multiple function
definitions when the target is unambiguous
and returns a diagnostic to the repair agent otherwise.

The given calling interface specifies the argument
and return types.
Using this information, the implementation defines
a common input layout for concrete execution and
symbolic analysis: how scalar arguments are represented,
how much memory is allocated for pointer arguments,
and which memory values are collected as outputs.
Buffer sizes are determined from the argument types,
configured size limits, and buffer lengths available
in the I/O test data.
A function may produce outputs by writing through
pointer arguments as well as returning a value.
The input layout identifies the buffers to collect
after execution, allowing these outputs to be
compared alongside the return value.
For concrete execution, the generated calling code
prepares arguments according to this layout,
invokes the function, and collects its outputs.
Symbolic execution uses the same layout to construct
symbolic arguments and buffers and to reconstruct
solver-generated inputs for concrete replay.

\subsection{Symbolic Execution and Solver Integration}
\label{sec:implementation-symbolic}

The verifier uses angr~\cite{shoshitaishvili2016state}
to explore the original and candidate binaries
independently.
Claripy, angr's symbolic-expression library,
represents symbolic inputs, path constraints,
and output expressions, with Z3~\cite{de2008z3}
as the primary solver.
The scheduler extends angr's \texttt{UniqueSearch},
retaining its execution-history similarity calculation.
It alternates between coverage-based selection,
which prioritizes states at unvisited basic blocks,
and the original history-based selection.
Traces from available tests initialize the
visited-block set, which is updated during exploration.
The scheduler also periodically selects the
longest-waiting state so that states not favored
by the coverage and history heuristics can
still make progress.
Recursive calls execute within the configured bounds,
while supported external functions are handled
through execution models.
The verifier retains completed-path constraints and
output expressions for comparison, together with
information about unfinished exploration and
execution failures.

For each pair of paths being compared, the verifier
asks the solver to find an input that satisfies
both path conditions and makes the observable
outputs differ.
Z3 is the default solver.
For comparisons involving floating-point arithmetic,
the verifier also uses
Bitwuzla~\cite{DBLP:conf/cav/NiemetzP23},
translating supported expressions into its native
floating-point operations.
If translation is unsupported or the backend
is unavailable, comparison falls back to
the primary Z3 solver path.
Separate time limits apply to symbolic exploration,
individual solver queries, and the overall verifier
invocation.
Memory and exploration bounds further limit
resource consumption.
The overall time limit covers exploration,
comparison, and concrete replay.
If analysis stops because a time, memory, or exploration
limit is reached, the verifier records the reason
and marks the assessment as incomplete.
Any counterexample confirmed by concrete replay
can still guide the repair agent's next revision.

\subsection{Counterexample Replay and Repair Integration}
\label{sec:implementation-feedback}

The verifier reconstructs counterexamples generated
by the symbolic comparison in
Section~\ref{sec:implementation-symbolic}
as concrete arguments using the shared input layout.
Argument types come from the given calling interface,
while argument values and input-buffer contents
come from the solver-generated counterexample.
Floating-point values retain their bit-level
representations when available.
The verifier then executes the original and recompiled
candidate binaries on these inputs and compares
their return values and values written to the
modeled output buffers.
The report indicates whether replay confirms the
difference and records the concrete inputs,
observed outputs, and values that differ.
Inputs and outputs are labeled using the types
specified by the calling interface, with exact
values retained for reproduction.

The controller validates and executes the repair
agent's requested actions.
After applying a repair, it repeats compilation
and testing following the workflow in
Section~\ref{sec:design-workflow}, returning the
results to the repair agent. %
Each assessment result is associated with the
candidate revision it checks.
The controller reuses that result only while the
candidate remains unchanged.
Before acceptance, it requires a symbolic assessment
of the current revision, so a result obtained
before a code change cannot justify accepting
the modified candidate.

Together, these components turn symbolic differences
into concrete, reproducible feedback for code repair.
Rechecking revised candidates allows the repair agent
to assess whether its changes resolve the reported
differences.

%% file: sections/06_evaluation_results.tex
\section{Evaluation}
\label{sec:results}

We evaluate \system{} with respect to three research questions:
\begin{itemize}
    \item \textbf{RQ1: Decompiled-Code Repair and Coverage.}
    How effectively does \system{} repair decompiled code,
    and how much coverage do its symbolically-generated inputs
    achieve across architectures and optimization levels?

    \item \textbf{RQ2: Contribution of Symbolic Feedback.}
    How does symbolic feedback affect single-step repair,
    and how effectively does symbolic comparison detect
    differences missed by the provided I/O tests?

    \item \textbf{RQ3: Analysis-Budget Sensitivity.}
    How does the analysis budget affect input coverage
    and symbolic comparison outcomes?
\end{itemize}

\subsection{Experimental Setup}
\label{sec:experimental-setup}

\paragraph{Datasets.}
We select 60 functions from ExeBench's %
\texttt{test\_real} split and 60 from DecompileBench.
To provide a consistent testing baseline, each task
uses ten concrete I/O tests, denoted IO10.
ExeBench provides these tests; for DecompileBench,
we generate inputs using angr~\cite{shoshitaishvili2016state}
and obtain expected outputs by executing the original binary.
We retain functions that compile for both x86-64
and ARM64 at O0--O3 and support replay of these tests,
yielding 960 binary decompilation tasks.
Each repair task starts from Ghidra's decompiled output.

\begin{figure*}[t]
    \centering
    \includegraphics[width=\textwidth]
        {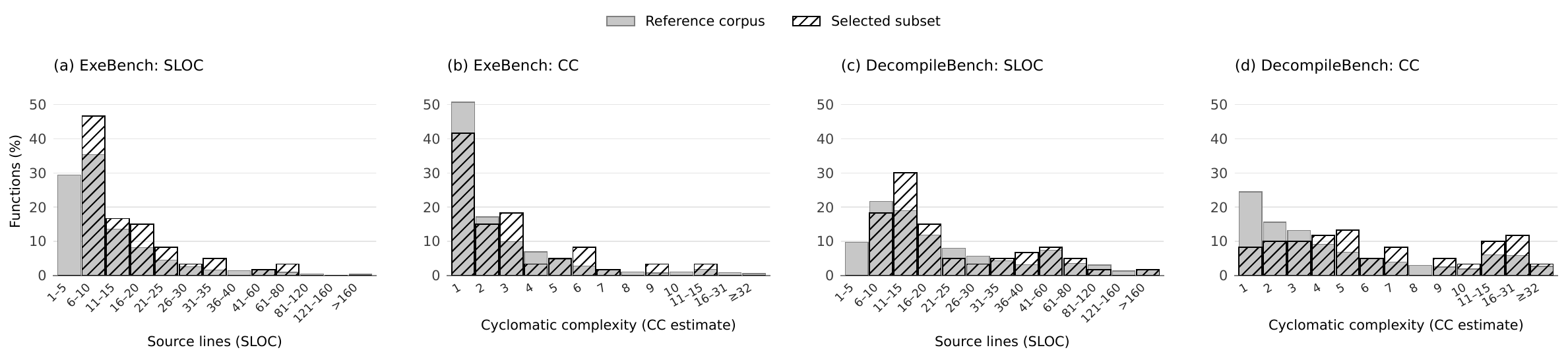}
    \caption{Source-level complexity distributions of
    the selected subsets and reference corpora.
    (a--b) ExeBench: source lines of code (SLOC) and
    estimated cyclomatic complexity (CC), comparing
    60 selected functions with the full
    \texttt{test\_real} split.
    (c--d) DecompileBench: the same metrics, comparing
    60 selected functions with the full corpus.
    Gray bars represent the reference corpus;
    hatched bars represent the selected subset.
    Bar heights show the percentage of functions
    within each group. CC is estimated lexically,
    and architecture and optimization variants are counted only once.}
    \label{fig:dataset-complexity}
\end{figure*}

Figure~\ref{fig:dataset-complexity} compares the
source-length and complexity distributions of the
selected subsets with ExeBench's full \texttt{test\_real}
split and the full DecompileBench corpus.
We measure source lines of code (SLOC) and a lexical
estimate of cyclomatic complexity (CC), counting
architecture and optimization variants only once.
The subsets contain fewer very short functions than
their reference corpora, and the DecompileBench subset
has a smaller proportion of functions with CC equal
to one.
Source snippets are deduplicated after removing
preprocessor directives and whitespace.
SLOC excludes blank lines, comments, and directives,
but includes function signatures and braces.

\paragraph{Environment and budgets.}
Our experimental platform is an Ubuntu 24.04.2 system with two AMD EPYC 9255
processors (48 physical cores) and 1~TiB of memory.
We use DeepSeek-V4-Flash via API,
angr for symbolic execution,
AFL++~\cite{fioraldi2020afl++} for fuzzing,
and QEMU for ARM64 concrete execution.
Symbolic comparison receives 60 seconds of exploration
per binary, while black-box fuzzing receives 120 seconds
for difference-detection input generation.
Symbolic comparison is required before acceptance;
confirmed differences and outstanding unverified
counterexamples block acceptance.
Symbolic exploration is limited to 800 paths and
12,000 steps, with an analysis memory limit of 24 GiB.
Coverage experiments use a 60-second generation budget;
RQ3 evaluates budgets of 30, 60, 120, 240,
and 3600 seconds.

\paragraph{LLM configuration.}
The repair agent uses \texttt{deepseek-v4-flash}
with provider-default temperature, top-$p$,
output-token limit, and thinking settings.
Each repair run is limited to 8 agent steps for
DecompileBench and 12 for ExeBench, with at most
3 patch attempts and 6 agent-requested probes.
The prompt includes the current candidate, verification
feedback, binary observations, and remaining budgets.
Complete prompt templates are provided in the
replication package.

\subsection{RQ1: Decompiled-Code Repair and Coverage}
\label{sec:rq1}

\textit{How effectively does \system{} repair decompiled
code, and what coverage do symbolically generated inputs
achieve across architectures and optimization levels?}

To answer this question, we evaluate final candidates
using compilation success, IO10 pass rates, and symbolic
comparison outcomes.
We also compare the line and branch coverage achieved
by symbolically generated inputs with IO10 and
the two fuzzing baselines.

\paragraph{Repair outcomes.}
Table~\ref{tab:optional-main} reports final outcomes
across datasets, architectures, and optimization levels,
averaged over three runs.
All percentages use all tasks as the denominator;
the symbolic outcome columns partition the candidates
passing IO10.
No diff denotes a completed comparison without
a detected difference under the modeled input domain,
while Partial no diff denotes an incomplete comparison
without a detected difference.
Diff denotes a replay-confirmed difference, including
findings from partial comparisons.

\begin{table*}[t]
\centering
\small
\setlength{\tabcolsep}{3.7pt}
\caption{Decompiled-code repair results (\%).
Ghidra denotes the initial decompiled code without
compatibility preprocessing or repair.
\system{} results are means over three runs;
each run contains 60 tasks per architecture--optimization
group and 480 per dataset.
Bold dataset rows report dataset-level results.
Compile: successful compilation and linking with
the test harness.
IO10: passes all ten fixed I/O tests.
All percentages use all tasks in the corresponding
group as the denominator.
The five symbolic-outcome columns partition the
IO10-passing candidates of \system{} and sum to
its IO10 column, up to rounding.
No diff / Partial: no behavioral difference detected
in a completed / incomplete comparison.
Diff: replay-confirmed difference, including findings
from partial comparisons.
``--'' denotes zero.}
\label{tab:optional-main}
\begin{tabular}{ll rr rr rrrrr}
\toprule
\multicolumn{2}{c}{Setting}
& \multicolumn{2}{c}{Ghidra}
& \multicolumn{2}{c}{\system{}}
& \multicolumn{5}{c}{Symbolic outcomes of \system{} IO10 passes} \\
\cmidrule(r){1-2}
\cmidrule(lr){3-4}
\cmidrule(lr){5-6}
\cmidrule(l){7-11}
Arch. & Opt.
& Compile & IO10
& Compile & IO10
& No diff & Partial & Diff & Timeout & Other \\
\midrule
\multicolumn{2}{l}{\textbf{DecompileBench (Avg.)}}
& \textbf{10.8} & \textbf{10.8}
& \textbf{99.0} & \textbf{98.8}
& \textbf{50.2} & \textbf{18.3}
& \textbf{0.5} & \textbf{28.1} & \textbf{1.8} \\
\addlinespace[2pt]
\multirow{4}{*}{\quad x86-64}
& O0 & 16.7 & 16.7 & 100.0 & 100.0
& 50.0 & 17.8 & -- & 29.4 & 2.8 \\
& O1 & 11.7 & 11.7 & 100.0 & 100.0
& 51.7 & 18.3 & -- & 28.3 & 1.7 \\
& O2 & 8.3 & 8.3 & 99.4 & 98.9
& 51.7 & 16.7 & 1.1 & 27.8 & 1.7 \\
& O3 & 8.3 & 8.3 & 98.9 & 98.9
& 51.7 & 19.4 & 0.6 & 25.6 & 1.7 \\
\addlinespace
\multirow{4}{*}{\quad ARM64}
& O0 & 11.7 & 11.7 & 98.3 & 98.3
& 47.8 & 12.8 & -- & 36.1 & 1.7 \\
& O1 & 11.7 & 11.7 & 99.4 & 98.9
& 50.0 & 21.1 & 0.6 & 25.6 & 1.7 \\
& O2 & 10.0 & 10.0 & 97.8 & 97.8
& 49.4 & 20.0 & 0.6 & 26.1 & 1.7 \\
& O3 & 8.3 & 8.3 & 97.8 & 97.8
& 49.4 & 20.0 & 1.1 & 25.6 & 1.7 \\
\midrule
\multicolumn{2}{l}{\textbf{ExeBench (Avg.)}}
& \textbf{18.8} & \textbf{16.0}
& \textbf{96.9} & \textbf{95.6}
& \textbf{62.7} & \textbf{14.4}
& \textbf{1.3} & \textbf{14.4} & \textbf{2.8} \\
\addlinespace[2pt]
\multirow{4}{*}{\quad x86-64}
& O0 & 18.3 & 8.3 & 98.3 & 95.0
& 64.4 & 12.8 & -- & 16.7 & 1.1 \\
& O1 & 23.3 & 23.3 & 97.8 & 97.2
& 65.6 & 12.8 & -- & 15.0 & 3.9 \\
& O2 & 18.3 & 16.7 & 98.9 & 98.3
& 64.4 & 15.6 & -- & 13.3 & 5.0 \\
& O3 & 16.7 & 15.0 & 95.0 & 95.0
& 61.7 & 16.1 & 0.6 & 15.0 & 1.7 \\
\addlinespace
\multirow{4}{*}{\quad ARM64}
& O0 & 16.7 & 11.7 & 97.8 & 97.8
& 65.6 & 15.0 & -- & 11.7 & 5.6 \\
& O1 & 18.3 & 18.3 & 95.6 & 95.6
& 63.9 & 12.2 & 1.1 & 16.1 & 2.2 \\
& O2 & 20.0 & 18.3 & 96.1 & 93.3
& 58.3 & 16.1 & 4.4 & 13.3 & 1.1 \\
& O3 & 18.3 & 16.7 & 95.6 & 92.2
& 57.8 & 14.4 & 4.4 & 13.9 & 1.7 \\
\bottomrule
\end{tabular}
\end{table*}

On DecompileBench, 99.0\% of final candidates compile
and 98.8\% pass IO10; the corresponding rates on
ExeBench are 96.9\% and 95.6\%.
Across architecture--optimization groups, IO10 pass
rates range from 97.8\% to 100.0\% on DecompileBench
and from 92.2\% to 98.3\% on ExeBench.
Of all tasks, 50.2\% on DecompileBench and 62.7\%
on ExeBench produce candidates that pass IO10 and
receive No diff.
A further 18.3\% and 14.4\%, respectively,
pass IO10 and receive Partial no diff.

At the end of the repair process, final candidates
for 0.5\% of DecompileBench tasks and 1.3\% of
ExeBench tasks pass IO10 but retain replay-confirmed
differences.
Although these discrepancies remain unresolved,
symbolic comparison provides concrete counterexamples
that can guide further automated repair or
manual inspection.

\paragraph{Input-generation coverage.}
Table~\ref{tab:coverage-960} compares target-function
line and branch coverage achieved by IO10, black-box
fuzzing, source-based fuzzing, and symbolic execution.
Generated inputs are replayed through a source-coverage
harness, counting only the target function and excluding
driver code. These measurements describe source coverage
achieved by the inputs, rather than instruction coverage
of the target binary.
Within each architecture--optimization group,
we compare all four methods on the same tasks with
valid coverage results and macro-average their
per-task coverage. The repair loop and coverage
experiment use the same symbolic exploration policy.

\begin{table*}[t]
\centering
\small
\caption{Target-function source coverage (\%).
IO10 denotes the ten fixed I/O tests;
BB, SF, and Sym denote black-box fuzzing,
source-based fuzzing, and symbolic execution.
BB, SF, and Sym use a 60-second input-generation
budget; IO10 is a fixed baseline.
Architecture and optimization labels identify the
binaries used for input generation.
All methods are evaluated on the same valid tasks
within each group, retaining 84.2\% of DecompileBench
tasks and 85.4\% of ExeBench tasks.
Dataset and overall averages pool the corresponding
valid tasks.
Bold and underlined values indicate the highest
and second-highest coverage in each row,
separately for line and branch coverage;
values tied after rounding are marked equally.}
\label{tab:coverage-960}
\begin{tabular*}{0.8\linewidth}
{@{\extracolsep{\fill}}llrrrr@{\hspace{12pt}}rrrr@{}}
\toprule
& &
\multicolumn{4}{c}{Line coverage}
& \multicolumn{4}{c}{Branch coverage} \\
\cmidrule(lr){3-6}
\cmidrule(lr){7-10}
Arch. & Opt.
& IO10 & BB & SF & Sym
& IO10 & BB & SF & Sym \\
\midrule
\multicolumn{2}{@{}l}{DecompileBench (Avg.)}
& 70.2 & 91.4 & \underline{92.9} & \textbf{93.5}
& 62.0 & 86.0 & \textbf{89.3} & \underline{88.9} \\
\cmidrule(lr){1-10}
x86-64 & O0
& 70.0 & \underline{93.2} & 92.3 & \textbf{95.1}
& 62.1 & \underline{89.7} & 89.1 & \textbf{92.7} \\
x86-64 & O1
& 70.6 & 92.1 & \underline{93.2} & \textbf{94.3}
& 62.6 & 86.6 & \textbf{89.7} & \underline{88.9} \\
x86-64 & O2
& 70.2 & 90.5 & \textbf{92.6} & \underline{92.4}
& 61.0 & 84.5 & \textbf{88.7} & \underline{87.3} \\
x86-64 & O3
& 70.1 & \underline{90.5} & \textbf{92.5} & \textbf{92.5}
& 61.8 & 84.7 & \textbf{88.7} & \underline{87.1} \\
\addlinespace[2pt]
ARM64 & O0
& 71.0 & \underline{93.2} & \underline{93.2} & \textbf{94.8}
& 62.6 & \underline{89.8} & \underline{89.8} & \textbf{91.9} \\
ARM64 & O1
& 69.2 & 91.5 & \underline{93.1} & \textbf{94.1}
& 60.7 & 85.7 & \textbf{89.6} & \underline{88.8} \\
ARM64 & O2
& 70.4 & 90.5 & \textbf{93.0} & \underline{92.5}
& 62.1 & 84.2 & \textbf{89.3} & \underline{87.1} \\
ARM64 & O3
& 70.4 & 89.6 & \textbf{93.5} & \underline{92.7}
& 62.8 & 82.9 & \textbf{89.8} & \underline{87.4} \\
\midrule
\multicolumn{2}{@{}l}{ExeBench (Avg.)}
& 93.1 & 98.1 & \underline{98.4} & \textbf{98.8}
& 87.3 & 95.4 & \textbf{97.3} & \underline{96.8} \\
\cmidrule(lr){1-10}
x86-64 & O0
& 92.7 & \underline{98.3} & \textbf{98.4} & \textbf{98.4}
& 86.3 & 97.1 & \textbf{97.4} & \underline{97.2} \\
x86-64 & O1
& 92.9 & \textbf{98.3} & \underline{98.2} & \textbf{98.3}
& 87.2 & 95.2 & \textbf{96.9} & \underline{96.5} \\
x86-64 & O2
& 93.3 & 97.8 & \textbf{98.5} & \underline{98.4}
& 88.0 & 94.1 & \textbf{97.5} & \underline{96.3} \\
x86-64 & O3
& 93.1 & 98.3 & \textbf{98.6} & \underline{98.4}
& 87.5 & 95.1 & \textbf{97.2} & \underline{96.1} \\
\addlinespace[2pt]
ARM64 & O0
& 93.4 & \underline{98.3} & \underline{98.3} & \textbf{98.8}
& 87.6 & \underline{97.0} & 96.8 & \textbf{98.1} \\
ARM64 & O1
& 92.9 & 97.8 & \underline{98.6} & \textbf{99.3}
& 87.2 & 94.4 & \textbf{97.4} & \underline{96.8} \\
ARM64 & O2
& 93.4 & 98.2 & \underline{98.5} & \textbf{99.3}
& 87.6 & 95.2 & \textbf{97.5} & \underline{96.8} \\
ARM64 & O3
& 93.0 & 97.9 & \underline{98.5} & \textbf{99.3}
& 86.8 & 95.0 & \textbf{97.3} & \underline{96.7} \\
\midrule
\multicolumn{2}{@{}l}{Overall}
& 81.8 & 94.8 & \underline{95.7} & \textbf{96.2}
& 74.7 & 90.7 & \textbf{93.3} & \underline{92.9} \\
\bottomrule
\end{tabular*}
\end{table*}

On DecompileBench, symbolic execution achieves 93.5\%
line coverage and 88.9\% branch coverage, compared
with 70.2\% and 62.0\% for IO10.
On ExeBench, it reaches 98.8\% and 96.8\%, compared
with 93.1\% and 87.3\% for IO10.
Symbolically generated inputs therefore achieve higher
average coverage than the supplied tests on both
datasets, with larger gains on DecompileBench.

Under the same 60-second input-generation budget,
symbolic execution also exceeds black-box fuzzing
in average line and branch coverage on both datasets.
Overall, it achieves 96.2\% line coverage and
92.9\% branch coverage, compared with 94.8\%
and 90.7\% for black-box fuzzing.
Symbolic execution achieves the highest overall
line coverage and approaches the branch coverage
of source-based fuzzing (92.9\% versus 93.3\%)
without using the reference source for input generation.
These results show that symbolic exploration exercises
more of the evaluated code than the supplied tests
and black-box fuzzing, providing broader coverage
for behavioral comparison.

\begin{rqanswer}{RQ1: Decompiled-Code Repair and Coverage}
\system{} produces candidates that pass IO10 for
97.2\% of the evaluated tasks.
For 56.5\% of all tasks, candidates pass the supplied
tests and complete symbolic comparison without
detected differences, providing stronger assurance
of behavioral consistency than test-suite agreement alone.
Under the same input-generation budget, symbolic
execution achieves higher average line and branch
coverage than black-box fuzzing on the same valid tasks,
supporting more effective exploration of program behavior.
\end{rqanswer}

\begin{figure}[t]
    \centering
    \includegraphics[width=0.6\linewidth]
        {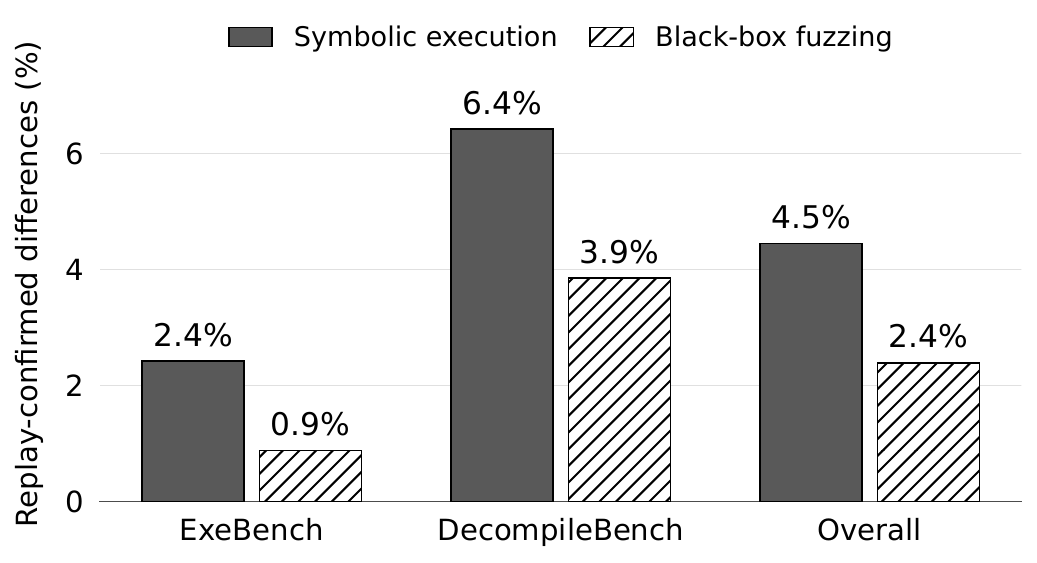}
    \caption{Replay-confirmed differences found by symbolic
    execution and black-box fuzzing.
    Both methods check the same candidates: the first
    candidate in each repair run to pass all ten fixed
    I/O tests.
    Each bar reports the percentage of these candidates
    with a confirmed behavioral difference.
    Confirmation requires an input that produces different
    outputs when executed on the original and recompiled
    binaries.}
    \label{fig:confirmed-differences}
\end{figure}

\subsection{RQ2: Contribution of Symbolic Feedback}
\label{sec:rq2}

\textit{How does symbolic feedback affect single-step
repair, and how effectively does symbolic comparison
detect differences missed by the provided I/O tests?}

To answer this question, we compare single-step repair
using IO replay alone, IO replay with symbolic feedback,
and IO replay with black-box fuzzing feedback.
We then evaluate how effectively symbolic execution
and black-box fuzzing detect behavioral differences
in candidates that already pass IO10.

\paragraph{Effect on single-step repair.}
Table~\ref{tab:single-step-deepseek-posthoc} compares
the three feedback groups from matched candidate
snapshots. For each eligible repair run, we retain
the first candidate that compiles but fails at least
one IO10 test, with at most one snapshot per task.
All groups receive IO replay feedback;
+Sym and +Fuzz additionally receive symbolic feedback
and black-box fuzzing feedback, respectively.
Feedback is collected before the model proposes a patch.
Each group then makes one repair attempt, without
additional agent-requested analysis or a second patch.
The resulting candidate is recompiled and evaluated.
We report IO10 pass rates over three repetitions
and offline DeepSeek assessments with counterexample
replay.
DeepSeek receives the reference and candidate source
code, function interface, and target architecture;
repair-group labels and prior detector outcomes are
omitted. This source-based assessment is performed
offline and is not provided to the repair agent.
Judged equivalent denotes a model judgment. Confirmed
diff requires a stable difference reproduced against
the original binary and reference source, with the
source sanitizer check passing; floating-point
differences are included. Other includes unconfirmed
counterexamples, invalid inputs, replay failures, and
source--binary disagreements.

\begin{table*}[t]
\centering
\small
\setlength{\tabcolsep}{4pt}
\caption{Single-step decompiled-code repair and offline
source-code assessment by DeepSeek.
All repair groups receive I/O test feedback; +Sym and +Fuzz
additionally receive symbolic and black-box fuzzing feedback,
respectively. IO10 pass rates are means $\pm$ sample standard
deviations over three repetitions; $\Delta$ is the improvement
over IO replay in percentage points.
In the assessment, DeepSeek compares the reference and candidate
source code: \emph{Judged equivalent} is a model judgment,
\emph{Confirmed diff} requires a counterexample confirmed by
concrete execution, and \emph{Other} is inconclusive.
Empty-response and output-format failures are excluded.
$\uparrow$/$\downarrow$: higher/lower is better;
the best result within each dataset is shown in bold.}
\label{tab:single-step-deepseek-posthoc}
\begin{tabular*}{\linewidth}{@{}@{\extracolsep{\fill}}llrrrrr@{}}
\toprule
& & \multicolumn{2}{c}{Repair}
& \multicolumn{3}{c}{DeepSeek assessment (\%)} \\
\cmidrule(lr){3-4} \cmidrule(lr){5-7}
Dataset & Group
& IO10 pass (\%) $\uparrow$ & $\Delta$ $\uparrow$
& Judged equiv.\ $\uparrow$ & Confirmed diff $\downarrow$ & Other $\downarrow$ \\
\midrule
\multirow{3}{*}{DecompileBench}
& IO replay & $97.0 \pm 3.0$ & -- & 84.5 & 11.5 & \textbf{4.0} \\
& +Sym (Ours) & $\mathbf{99.1 \pm 1.6}$ & $\mathbf{+2.1}$ & \textbf{91.4} & \textbf{4.6} & \textbf{4.0} \\
& +Fuzz & $97.6 \pm 3.0$ & $+0.6$ & 85.0 & 10.4 & 4.6 \\
\midrule
\multirow{3}{*}{ExeBench}
& IO replay & $86.4 \pm 4.0$ & -- & 78.2 & 11.2 & 10.6 \\
& +Sym (Ours) & $\mathbf{90.2 \pm 2.2}$ & $\mathbf{+3.8}$ & \textbf{82.5} & \textbf{7.3} & \textbf{10.3} \\
& +Fuzz & $87.1 \pm 0.8$ & $+0.7$ & 78.5 & 8.8 & 12.7 \\
\midrule
\multirow{3}{*}{Overall}
& IO replay & $90.1 \pm 1.7$ & -- & 80.4 & 11.3 & 8.3 \\
& +Sym (Ours) & $\mathbf{93.3 \pm 2.0}$ & $\mathbf{+3.2}$ & \textbf{85.5} & \textbf{6.3} & \textbf{8.1} \\
& +Fuzz & $90.7 \pm 0.6$ & $+0.6$ & 80.8 & 9.3 & 9.9 \\
\bottomrule
\end{tabular*}
\end{table*}

Across both datasets, +Sym achieves a mean IO10 pass
rate of 93.3\%, exceeding IO replay and +Fuzz by
3.2 and 2.6 percentage points, respectively.
+Sym and +Fuzz reach 99.1\% and 97.6\%, respectively,
on DecompileBench,
while +Sym achieves 90.2\% on ExeBench, compared
with 86.4\% for IO replay and 87.1\% for +Fuzz.
The offline assessments show a corresponding trend:
the pooled confirmed-difference rate is 6.3\%
for +Sym, compared with 11.3\% for IO replay and
9.3\% for +Fuzz.
Thus, in the reported ablation, symbolic feedback
combines the highest mean test-pass rate with the
lowest pooled rate of replay-confirmed differences.

\paragraph{Differences beyond IO10.}
We next examine the first IO10-passing candidate
from each repair run, whether it is the initial
candidate or a later revision.
Symbolic execution and black-box fuzzing independently
check the same candidates.
A difference is counted only when the method's own
counterexample is confirmed through concrete replay.
Detection rates use all selected candidates in each
dataset as the denominator.

Figure~\ref{fig:confirmed-differences} shows that
symbolic execution detects differences in 4.5\%
of the selected candidates overall, compared with
2.4\% for black-box fuzzing.
The respective rates are 2.4\% and 0.9\%
on ExeBench, and 6.4\% and 3.9\%
on DecompileBench.
Under the evaluated configuration, symbolic
execution detects differences in more IO10-passing
candidates on both datasets, exposing behavioral
discrepancies missed by the provided tests.

\begin{rqanswer}{RQ2: Contribution of Symbolic Feedback}
Symbolic feedback delivers the highest mean
single-step IO10 pass rate in our ablation,
exceeding IO replay alone and additional fuzzing
feedback by 3.2 and 2.6 percentage points,
respectively.
Beyond IO10, symbolic comparison exposes
replay-confirmed differences in 4.5\% of the
same test-passing candidates, compared with
2.4\% for black-box fuzzing.
These results show that symbolic comparison both
detects behavioral differences missed by the supplied
tests and provides useful feedback for repairing
decompiled code.
\end{rqanswer}

\subsection{RQ3: Analysis-Budget Sensitivity}
\label{sec:rq3}

\textit{How does the analysis budget affect input
coverage and symbolic comparison outcomes?}

To answer this question, we examine coverage under
different input-generation budgets and comparison
outcomes under different total time budgets.

\paragraph{Input-generation coverage.}
We evaluate symbolic execution,
black-box fuzzing, and source-based fuzzing on the
selected ExeBench functions with budgets of 30, 60,
120, 240, and 3600 seconds per function.
We report line coverage, branch coverage, and the
fraction of functions achieving 100\% coverage on
both metrics.
The functions and each method's initialization policy
remain fixed across budgets; IO10 serves as a
fixed reference.

\begin{figure}[t]
    \centering
    \includegraphics[width=\linewidth]
        {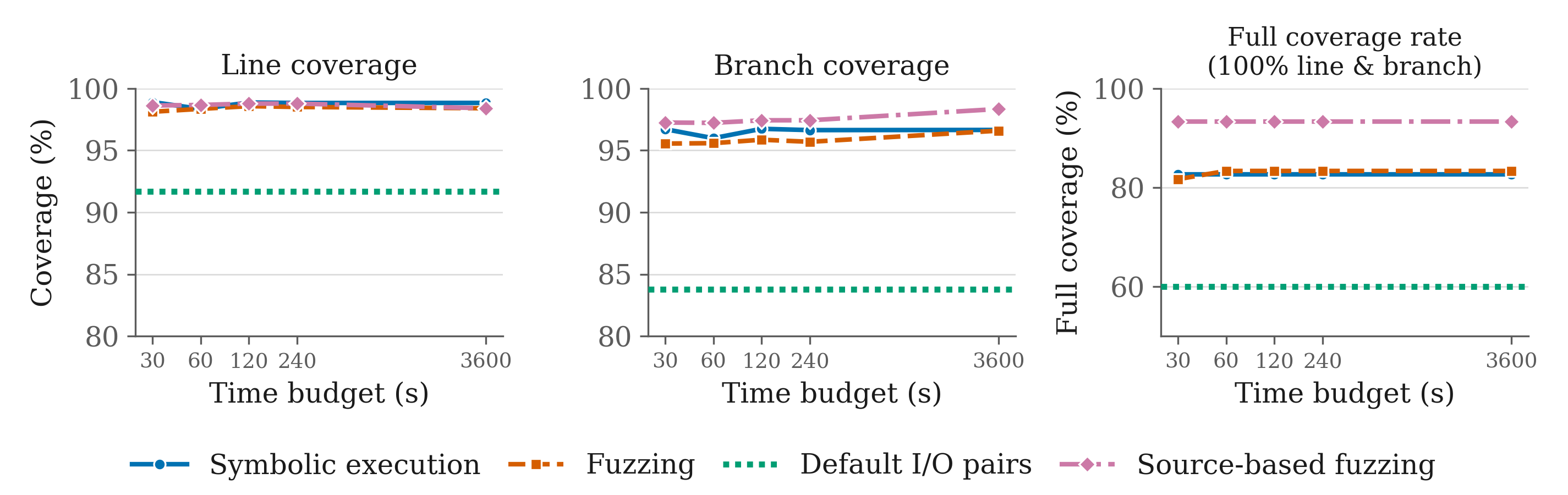}
    \caption{ExeBench coverage under input-generation
    budgets of 30, 60, 120, 240, and 3600 seconds.
    Methods are compared at the same budget.
    IO10 uses the ten supplied I/O tests as a fixed baseline.
    Line and branch coverage measure the code exercised
    within each target function.
    Full coverage reports the percentage of valid cases
    with both 100\% line and 100\% branch coverage.}
    \label{fig:coverage_time}
\end{figure}

Figure~\ref{fig:coverage_time} shows that symbolic
execution achieves similar line and branch coverage
across budgets of 30--240 seconds, with little further
change at 3600 seconds.
Symbolically-generated inputs achieve higher coverage
than IO10 throughout the sweep.
Compared with black-box fuzzing, symbolic execution
achieves higher branch coverage at the shorter budgets,
but the gap narrows at 3600 seconds.
Source-based fuzzing achieves a higher full-coverage
rate across the tested budgets and increases its
branch coverage at 3600 seconds.
These observations support 60 seconds as a practical
budget for the evaluated ExeBench configuration,
although they do not establish it as an optimal budget.

\begin{figure*}[t]
    \centering
    \includegraphics[width=\linewidth]
        {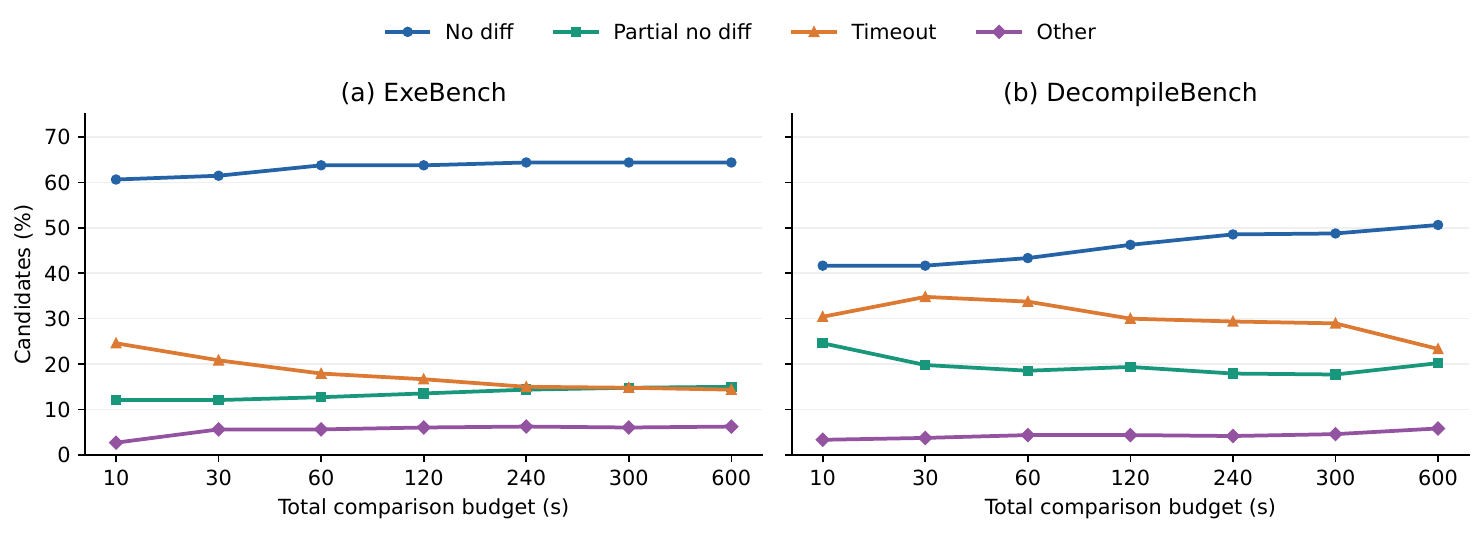}
    \caption{Symbolic comparison outcomes under different
    total time budgets on (a) ExeBench and
    (b) DecompileBench.
    Each dataset contains 480 candidates held fixed
    across budgets.
    Timeout includes both inner and outer timeouts.
    Other includes detected differences and all remaining
    outcomes.
    The four categories sum to 100\% at each budget.
    Budget levels are equally spaced.}
    \label{fig:symbolic-budget}
\end{figure*}

\paragraph{Symbolic comparison outcomes.}
We also examine how the total time budget affects symbolic
comparison outcomes on 480 fixed candidates from %
each dataset, one per binary decompilation task. The candidates remain unchanged
across budgets, allowing us to isolate the effect
of additional analysis time.
Figure~\ref{fig:symbolic-budget} reports the
percentages of candidates classified as No diff,
Partial no diff, Timeout, or Other.
No diff indicates that the configured comparison
completed without finding a difference.
Partial no diff indicates that no difference was
found in the completed portion of the analysis.
Timeout includes both inner and outer timeouts;
Other groups detected differences and all
remaining outcomes.

\begin{rqanswer}{RQ3: Analysis-Budget Sensitivity}
Symbolic input generation reaches a coverage plateau
within the shorter tested budgets on the evaluated
ExeBench functions. Extending the budget to 3600 seconds
provides little additional symbolic coverage, although
the gap with black-box fuzzing narrows.
The comparison-outcome sweep separately shows how
completed and partial assessments vary with total
analysis time on fixed candidates.
\end{rqanswer}

%% file: sections/07_discussion_limitations.tex
\section{Discussion and Limitations}
\label{sec:limitations}

This section discusses the scope and limitations
of symbolic comparison and the factors that may
affect the validity of our evaluation.

\subsection{Scope and Limitations of Verification}

The verifier requires the target function's calling
interface, including argument and return types,
to construct symbolic inputs and interpret outputs.
In our experiments, this information is provided
with the evaluation tasks, together with ten I/O
pairs per task. In other settings, the interface
can be supplied by a user or recovered through
binary analysis.

Available concrete tests help prioritize exploration.
With a configured input model, symbolic execution can
also start without seed tests; its inputs are not
restricted to the values in the given tests.
However, the search remains bounded by the modeled
input domain and available analysis time, so it may
not expose every behavioral difference.

The verifier compares return values and modeled
parameter-memory outputs using finite buffers.
Supported external functions are handled through
models of their behavior. This allows functions
to be analyzed without reconstructing the entire
application, but the comparison remains limited
to the behavior represented by these models.
Functions that depend on files, global state,
concurrency, or kernel interaction require additional
environment support. Likewise, finite buffers do
not cover every possible memory layout or object size.
Within these bounds, our evaluation shows that symbolic
execution achieves higher average line and branch
coverage than black-box fuzzing and exposes
replay-confirmed differences in candidates that pass
all supplied tests (Sections~\ref{sec:rq1}
and~\ref{sec:rq2}).

Path explosion limits the scalability of symbolic
execution~\cite{yao2025empc}, so our analysis operates
within a fixed resource budget.
Although exploration and comparison may remain
incomplete, explored paths can still reveal behavioral
differences. Once concrete replay confirms a
counterexample, the repair agent can use it to guide
repair without waiting for exhaustive analysis.

About one-fifth of the evaluated functions contain
floating-point arithmetic or comparisons.
The verifier uses Bitwuzla to reason about
floating-point operations and rounding.
Binary code can also combine floating-point arithmetic
with operations on the underlying bit representation,
such as modifying exponent or mantissa bits before
reinterpreting the result as a floating-point value.
When our translation layer cannot handle these
expressions, or Bitwuzla is unavailable, the verifier
falls back to the default Z3-based comparison.
However, Z3 may struggle with these mixed constraints
and reach its time or resource limits, leaving the
comparison unresolved.

Concrete replay confirms that a counterexample
produces different outputs on the original and
recompiled binaries, reducing the risk of reporting
differences caused by symbolic modeling errors.
Some counterexamples may involve inputs that do
not occur in normal application use.
Application-specific input constraints can help
focus the comparison on behavior relevant to
the intended deployment.
When no difference is found, the strength of that
result depends on the execution model and whether
exploration and comparison have completed.
In our evaluation, 56.5\% of all tasks produce
candidates that pass the supplied tests and complete
symbolic comparison without detected differences,
providing stronger evidence of behavioral consistency
than test-suite agreement alone
(Section~\ref{sec:rq1}).

\subsection{Threats to Evaluation Validity}

LLMs can produce different patches for the same
candidate, causing repair outcomes to vary.
We repeat the repair experiments three times
to reduce the influence of this variation.
In the paired ablation, all feedback groups start
from the same candidate and use consistent model
settings and patch-processing rules.
As an offline assessor, DeepSeek may also overlook
behavioral differences or propose invalid counterexamples,
even when given both the reference and candidate source code.
We check DeepSeek's proposed counterexamples
by executing them.
If execution does not reproduce the claimed
difference, the assessment remains inconclusive;
it is not counted as a confirmed difference
or evidence of equivalence.

We select functions that compile for both x86-64
and ARM64 and support ten replayable I/O pairs,
allowing consistent comparisons across architectures
and optimization levels. This selection may exclude
functions that depend on external calls or execution
environments unsupported by our setup. To examine
whether the selection favors smaller or simpler
functions, we compare the selected subsets with
their reference corpora in source size and complexity.

%% file: sections/08_related_work.tex
\section{Related Work}
\label{sec:related-work}

\paragraph{LLM-based decompilation.}
LLM4Decompile develops specialized models for
source recovery and decompiler-output refinement
\cite{tan2024llm4decompile}.
DeGPT uses specialized LLM roles to improve
decompiled code \cite{hu2024degpt}, while
CodeInverter incorporates control-flow and
data-mapping information into source recovery
\cite{liu2025codeinverter}.
Beyond generation, recent work uses analysis
and execution feedback to improve code quality~\cite{su2025disco,ibrahimzada2025alphatrans}.
D-LiFT uses compilation and symbolic checks
in a quality assessment that guides model
training \cite{zou2025d}.
MCGD and AutoDecompiler incorporate validation
feedback into repeated generation attempts
\cite{zhang2026mcgd,liu2026autodecompiler}.
CHISEL further uses the original binary as an
executable reference, combining compiler diagnostics
with type-aware differential fuzzing
\cite{chisel2026}.
We share this use of a binary reference, but focus
on symbolic counterexamples as feedback for
LLM-based source repair and evaluate their effect
against I/O replay and fuzzing feedback.

\paragraph{Symbolic comparison and decompiler validation.}
Differential Symbolic Execution characterizes
behavioral changes between program versions
\cite{person2008dse}.
SymDiff checks equivalence over Boogie
representations \cite{lahiri2012symdiff},
while ARDiff improves scalability by abstracting
and refining common code \cite{ardiff2020}.
Related approaches compare executable behavior:
BinHunt combines control-flow matching with
symbolic reasoning for binary comparison
\cite{gao2008binhunt}, and UC-KLEE and KLEE-CL
cross-check function implementations and
data-parallel floating-point code, respectively
\cite{ramos2011practical,collingbourne2014symbolic}.
D-Helix applies symbolic comparison directly
to original and recompiled binaries and adjusts
decompiler heuristics to correct detected errors
\cite{zou2024dhelix}.
Checking the output of a transformation against
its input also relates to translation validation
\cite{pnueli1998translation}, including Alive2's
bounded validation of LLVM optimizations
\cite{lopes2021alive2}.
Our system performs comparison on native binaries
through angr \cite{shoshitaishvili2016state}.
Unlike D-Helix's heuristic adjustment, it returns
replay-confirmed counterexamples to an LLM that
revises the candidate source.

\paragraph{Symbolic test generation and hybrid fuzzing.}
Symbolic and concolic testing use execution
constraints to generate inputs beyond those
already exercised.
CUTE combines concrete and symbolic execution
for testing C programs \cite{sen2005cute},
and KLEE generates high-coverage tests for
systems programs \cite{cadar2008klee}.
SAGE solves constraints collected from concrete
traces to generate inputs for alternative paths
\cite{godefroid2008automated}.
Hybrid fuzzing combines this reasoning with
mutation-based exploration: Driller uses selective
concolic execution to complement fuzzing
\cite{stephens2016driller}, while QSYM develops
an efficient concolic engine for this setting
\cite{yun2018qsym}.
Our exploration policy also uses concrete
execution information to guide symbolic search.
Its purpose is to support comparison between
an original binary and a recompiled candidate,
so that newly exposed behavior can provide
counterexamples for source repair.

\paragraph{Feedback-guided program repair.}
Conversational program repair incorporates
previous patches and validation results into
subsequent model requests
\cite{xia2023conversational}, while Repilot
uses completion-engine feedback to improve
patch validity \cite{wei2023repilot}.
ReAct provides a broader interaction pattern
in which an LLM selects actions and uses their
observations to guide subsequent decisions
\cite{yao2022react}.
Our agent follows this pattern, with binary
analysis and symbolic comparison available
as tools.
Symbolic repair systems use semantic information
more directly in patch construction:
SemFix derives repair constraints
\cite{nguyen2013semfix}, DirectFix seeks simple
repairs through constraint solving
\cite{mechtaev2015directfix}, and Angelix and S3
use semantic guidance for patch synthesis
\cite{mechtaev2016angelix,le2017s3}.
Fix2Fit additionally connects patch search with
test generation to distinguish candidate patches
\cite{gao2019crash}.
Our approach follows a counterexample-guided
process \cite{solarlezama2006sketch}, but assigns
different roles to analysis and generation:
symbolic comparison identifies behavioral
differences, while the LLM revises the source.

\paragraph{Repair correctness and evaluation.}
Passing the available tests does not establish
patch correctness, since a repair can overfit
those tests \cite{smith2015overfitting}.
Patch-correctness studies therefore examine
evidence beyond the original test suite
\cite{xiong2018identifying,yang2023patchcorrectness}.
However, additional generated tests do not
necessarily turn incorrect patches into correct
ones \cite{yu2017test}.
This distinction motivates our separate
evaluations of difference detection and repair
effectiveness.
In decompilation, the goal of correction also
depends on how successful recovery is defined.
Decomperson studies source recovery aimed at
reproducing the original binary through compilation
\cite{burk2022decomperson}, while Walking the
Last Mile examines how practitioners correct
Python decompiler output
\cite{wiedemeier2025lastmile}.
Our setting concerns behavioral agreement of
native binary functions rather than exact binary
reproduction, and we evaluate whether counterexample
feedback helps an LLM achieve that agreement.

%% file: sections/09_conclusion.tex
\section{Conclusion}

We presented \system{}, an LLM-assisted decompiled-code
repair system that uses symbolic comparison to
assess behavioral consistency and guide code revisions
with replay-confirmed counterexamples.
Across 960 binary decompilation tasks from ExeBench and DecompileBench,
symbolically generated inputs achieve higher coverage
than black-box fuzzing under the same input-generation
budget. More than half of the tasks yield candidates
that pass the provided tests and complete symbolic
comparison with no behavioral difference found.
In a paired single-step experiment, symbolic feedback
improves the mean test-pass rate by 3.2 percentage
points over test feedback alone.
These results demonstrate the value of symbolic
comparison for assessing and repairing decompiled
code beyond fixed tests.
Future work will extend support for external functions
and complex memory behavior and improve exploration
under practical resource budgets.

%% file: sections/references.tex
\bibliographystyle{ACM-Reference-Format}
\bibliography{references}